\documentclass[twocolumn,aps,prl,showpacs]{revtex4}

\usepackage{latexsym,amssymb,float}
\usepackage{setspace}
\usepackage{graphicx}
\usepackage{epsfig}

\newcommand{\e}{\varepsilon}

\newcommand{\f}{\frac}

\begin{document}

\title{Multiband Superconductivity in Spin Density Wave Metals}
\author{J.-P. Ismer$^{1,2}$, Ilya Eremin$^{1,2}$, Enrico
Rossi$^{3,*}$,
Dirk K. Morr$^{3}$, G. Blumberg$^{4}$}
\affiliation{$^1$Max-Planck-Institut f\"ur Physik komplexer Systeme, 01187 Dresden, Germany \\
$^2$Institut f\"ur Mathematische Physik,
TU Braunschweig, 38106
Braunschweig, Germany\\
$^3$Department of Physics, University of Illinois at
Chicago, Chicago, IL 60607, USA\\
$^4$Department of Physics and Astronomy, The State University of
New Jersey, Piscataway, NJ 08854, USA}
\date{\today}

\begin{abstract}
We study the emergence of multiband superconductivity with $s$- and
$d-$wave symmetry on the background of spin density wave (SDW). We
show that the SDW coherence factors renormalize the momentum
dependence of the superconducting (SC) gap, yielding a SC state with
an \emph{unconventional} $s$-wave symmetry. Interband Cooper pair
scattering stabilizes superconductivity in both symmetries. With
increasing SDW order, the $s$-wave state is more strongly suppressed
than the $d$-wave state. Our results are universally applicable to
two-dimensional systems with a commensurate SDW.

\end{abstract}

 \maketitle

Understanding the microscopic origin of thermodynamic phases with
multiple order parameters (OP) is one of the central issues in
condensed matter physics. This topic is particularly important for
describing the complex phase diagram of many correlated metals such
as the high-T$_c$ cuprates, ferropnictides, and heavy fermion
compounds, in which it was suggested that superconductivity coexists
with a rich variety of other states, such as charge, spin, or
orbital density wave states. Interestingly enough, in such
coexistence phases, the presence of a density wave immediately leads
to multiband superconductivity due to the folding of the electronic
bands. The relative phase of the superconducting (SC) order
parameters in multiband superconductivity was originally
studied in Ref.~\cite{Legget}, and has attracted significant
interest recently in the context of MgB$_2$ \cite{blumberg_0} and
the ferropnictides \cite{mazin}, in which in-phase and out-of-phase
locking, respectively, of the OPs has been found.

The growing experimental evidence for the coexistence of
superconductivity and a spin density wave (SDW) in the \emph{n}-type
(i.e., electron-doped) cuprates \cite{review,greene,matsui}, some
heavy fermion \cite{pfleiderer} and organic superconductors
\cite{lebed} raises the question of how phase-locking
occurs in such systems. This question is of particular interest in
the \emph{n}-type cuprates due to a possible transition of the SC
symmetry from $d$-wave to $s$-wave with increasing doping. However,
there remains experimental disagreement regarding this transition.
While some measurements are consistent with a transition from a SC
$d_{x^2-y^2}$-wave symmetry in the underdoped materials to either an
$s$- or a ($d+is$)-wave symmetry in the optimally and overdoped
ones, other results suggest the existence of a $d_{x^2-y^2}$-wave OP
with higher harmonics for the entire doping range \cite{review}.

In this Letter, we study the emergence of two-band superconductivity
with $d_{x^2-y^2}$- or $s$-wave symmetry on the background of a
commensurate SDW state with imperfect nesting, using the
\emph{n}-type cuprates  as a particular example. We show that the
SDW coherence factors renormalize the momentum dependence of the SC
gap. This yields an $s$-wave OP which is \emph{unconventional}, in
that it acquires a $\pi$-phase shift between the two bands, line
nodes along the boundary of the reduced Brillouin zone (RBZ), and
changes sign between momenta connected by the SDW ordering momentum
${\bf Q}$. In contrast, in the $d_{x^2-y^2}$-wave state, the OP is
locked in-phase, with no {\it additional} line nodes. In both cases,
superconductivity is stabilized by interband Cooper pair scattering. With
increasing SDW OP, the \emph{unconventional} $s$-wave state is
suppressed more quickly than the $d_{x^2-y^2}$-wave state. In the
latter case, the vanishing of $T_c$ coincides approximately with the
disappearance of the first of the two Fermi surfaces (FS). To
demonstrate the generality of our results for other materials, we
consider both $n$- and $p$-type doping, leading to different
evolutions of the FS in the SDW state.

\begin{figure}[!t]
\includegraphics[width=8.2cm]{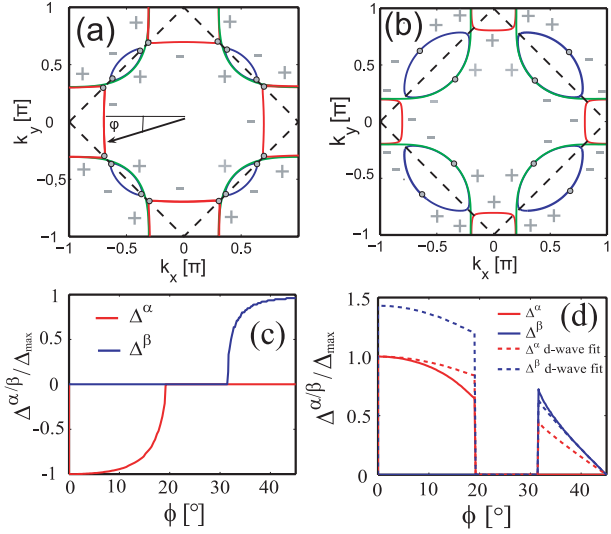}
\caption{(color online) FS (green lines) for the \emph{n}-type (a)
and \emph{p}-type (b) cuprates ($x=0.12$) in the paramagnetic state.
In the SDW state ($W_0 =0.1$~eV), the $\alpha$-band possesses
electron-pockets (red) and the $\beta$-band hole pockets (blue). The
RBZ is shown as a dashed line. The phase of the (a)
\emph{unconventional} $s$ and (b) $d_{x^2-y^2}$ SC OPs are denoted
by $+/-$. (c), (d) Angular dependence of the SC OP in the
\emph{unconventional} $s$-wave and $d_{x^2-y^2}$-wave channels for
\emph{n}-type doping close to $T_c$.} \label{fig:Fermisurf}
\end{figure}

Our starting point for investigating the coexistence of
superconductivity and SDW order is the Hamiltonian
\begin{eqnarray}
\mathcal{H} & = & \sum_{\bf k\sigma} \varepsilon_{\bf k}
c^{\dagger}_{\bf k \sigma} c_{\bf k \sigma} - \sum_{\bf
k,k^\prime,\sigma} U c_{{\bf k}\sigma}^{\dagger} c_{{\bf k+Q}\sigma}
c_{{\bf k^\prime+Q} \bar{\sigma}}^{\dagger} c_{{\bf
k^\prime}\bar{\sigma}}\nonumber
\\&&+\sum_{{\bf k,p,q,\sigma}}V_{\bf
q}\, c_{{\bf k+q}\sigma}^{\dag}c_{{\bf
p-q}\bar{\sigma}}^{\dag}c_{{\bf p}\bar{\sigma}}c_{{\bf k}\sigma}
\label{eq:1}
\end{eqnarray}
where $c_{\bf k\sigma}^{\dag}$ ($c_{\bf k\sigma}$) creates
(annihilates) an electron with spin $\sigma$ and  momentum ${\bf
k}$. We consider a two-dimensional system with the normal state
tight-binding energy dispersion $\varepsilon_{\bf k}=-2t\left( \cos
k_x + \cos k_y \right) + 4 t^\prime \cos k_x \cos k_y -2t^{\prime
\prime} \left(\cos 2 k_x + \cos 2 k_y \right)- \mu$ and hopping
matrix elements $t = 250$ meV, $t^\prime/t = 0.4$, and $t^{\prime
\prime}/t = 0.1$. The chemical potentials $\mu/t = -0.32$ and $\mu/t
= -1.09$ describe the slightly underdoped \emph{n}- and
\emph{p}-type cuprates, respectively, and the corresponding FS in
the paramagnetic state are shown in Figs.~\ref{fig:Fermisurf}\,(a)
and (b). The second and third terms in Eq.~(\ref{eq:1}) give rise to
a commensurate SDW and superconductivity, respectively. While it is
generally assumed that both phases emerge from the same underlying
interaction, its renormalization due to vertex corrections gives
rise to two different effective interactions, $U$ and $V_q$, assumed
to be independent of each other.

\begin{figure}[!t]
\includegraphics[width=8.25cm]{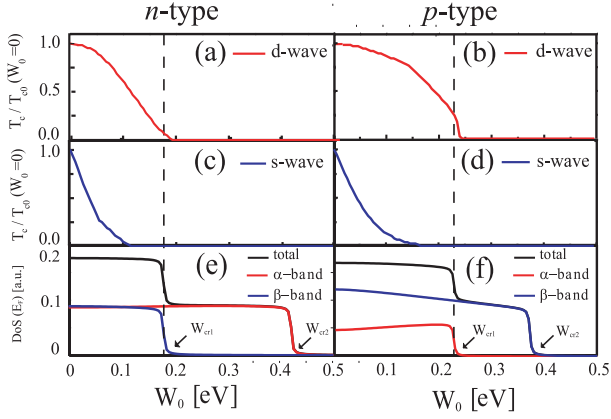}
\caption{(color online) $T_c$ for \emph{unconventional} $s$- and
$d$-wave symmetries, as well as the DOS, as a function of $W_0$ for
(a), (c), (e) \emph{n}- and (b), (d), (f) \emph{p}-type cuprates. We
set $W_0 =0.1$~eV, and for the $n$-type ($p$-type) cuprates $V_d=2$
eV ($V_d=1.2$ eV) and $V_s=0.6$ eV ($V_s=0.47$ eV).} \label{fig2}
\end{figure}

In what follows, we study the coexistence phase in the limit $T_{c}
\ll T_{SDW}$, and diagonalize the Hamiltonian in two steps. We first
derive the electronic spectrum of the SDW state by decoupling the
second term of Eq.(\ref{eq:1}) \emph{via} a mean-field (MF)
approximation and then diagonalize it together with the first term
using an SDW Bogoliubov transformation, introducing new
quasi-particle operators $\alpha_{\bf k}, \beta_{\bf k}$ for the two
resulting bands with dispersions
$E_{\bf k}^{{\alpha,\beta}} = \e^+_{\bf k}  \pm
\sqrt{\left(\e^-_{\bf k}\right)^{2}+W_0^{2}}$.
%\end{eqnarray}
%
Here, $W_0= U/2 \sum_{{\bf k},\sigma}
  \langle c^{\dagger}_{\bf k+Q, \sigma} c_{\bf
k, \sigma} \rangle {\rm sgn}\sigma$ is the SDW order parameter with
ordering momentum ${\bf Q}=(\pi,\pi)$, and $\varepsilon^\pm_{\bf
k}=\left( \varepsilon_{\bf k} \pm \varepsilon_{\bf k+Q} \right)/2$.
Due to imperfect nesting the $\alpha$ band exhibits electron pockets
around $(\pm \pi, 0)$ and $(0,\pm \pi)$ while the $\beta$ band
possesses hole-pockets around $(\pm \pi/2, \pm \pi/2)$, as shown in
Figs.~\ref{fig:Fermisurf}(a) and (b) for the \emph{n}- and
\emph{p}-type cuprates, respectively.  The size of these pockets
decreases with increasing $W_0$. At $W_0=W_{cr1} \approx 0.19$ eV
($W_{cr1} \approx 0.23$ eV), the hole (electron) pockets disappear
first for the {\it n}-type ($p$-type) cuprates, followed by the
vanishing of the electron (hole) pockets at $W_{cr2}>W_{cr1}$, as
shown by the evolution of the density of states (DOS) in
Figs.~\ref{fig2}(e) and (f). For $W>W_{cr2}$, the system is an
antiferromagnetic insulator (AFI).

We next apply the SDW Bogoliubov transformation to the SC pairing
interaction [third term in Eq.(\ref{eq:1})], and subsequently
performing a MF decoupling in the particle-particle channel, keeping
only anomalous expectation values of the form $\langle
\alpha^\dagger_{{\bf k}, \uparrow} \alpha^\dagger_{-{\bf k},
\downarrow} \rangle$ and $\langle \beta^\dagger_{{\bf k}, \uparrow}
\beta^\dagger_{-{\bf k}, \downarrow} \rangle$, and their complex
conjugates. The resulting MF Hamiltonian can be diagonalized by two
independent Bogoliubov transformations, yielding $\Omega_{{\bf
k}}^{\gamma} = \sqrt{\left(E_{{\bf
k}}^{\gamma}\right)^{2}+\left(\Delta_{{\bf k}}^{\gamma}\right)^{2}}$
($\gamma = \alpha, \beta$) as the energy dispersion of the two
bands. The SC gaps, $\Delta_{{\bf k}}^{\alpha,\beta}$ are determined
self-consistently from two coupled gap equations given by (at T =
0\,K)
%\begin{widetext}
\begin{eqnarray}
\Delta_{{\bf k}}^{\alpha} & = & -\sum_{{\bf p}\in RBZ}\left[
L^{\alpha \alpha}_{\bf k,p} \f{\Delta_{{\bf
p}}^{\alpha}}{2\Omega_{{\bf p}}^{\alpha}}  + L^{\alpha \beta}_{\bf
k,p}\f{\Delta_{{\bf p}}^{\beta}}{2\Omega_{{\bf
p}}^{\beta}}  \right] \nonumber\\
\Delta_{{\bf k}}^{\beta} & = & -\sum_{{\bf p}\in RBZ}\left[ L^{\beta
\alpha}_{\bf k,p} \f{\Delta_{{\bf p}}^{\alpha}}{2\Omega_{{\bf
p}}^{\alpha}}  + L^{\beta \beta}_{\bf k,p}\f{\Delta_{{\bf
p}}^{\beta}}{2\Omega_{{\bf
p}}^{\beta}}  \right] \label{eq:gaps}
\end{eqnarray}
%\end{widetext}
where $L^{\alpha \alpha}_{\bf k,p} = L^{\beta \beta}_{\bf k,p}
=\left( V_{\bf k-p} F^{u,v}_{{\bf k},{\bf p}} - V_{\bf k-p+Q}
F^{v,u}_{{\bf k},{\bf p}} \right)$, $L^{\alpha \beta}_{\bf k,p} =
L^{\beta \alpha}_{\bf k,p} = \left( V_{\bf k-p}N^{v,u}_{{\bf k},{\bf
p}} - V_{\bf k-p+Q} N^{u,v}_{{\bf k},{\bf p}} \right)$ with
$N^{x,y}_{{\bf k},{\bf p}}, F^{x,y}_{{\bf k},{\bf p}}, =u^2_{\bf
k}x^2_{\bf p} \pm 2u_{\bf k}v_{\bf k}u_{\bf p}v_{\bf p} + v^2_{\bf
k}y^2_{\bf p}$, $x,y=u,v$, $u_{\bf k}^{2}, v_{\bf k}^{2} =
\frac{1}{2} \left[ 1 \pm \frac{\varepsilon^-_{\bf
k}}{\sqrt{\left(\varepsilon^-_{\bf k}\right)^{2}+W_0^{2}}} \right]$,
and $u_{\bf k} v_{\bf k}  =
\frac{W_0}{2\sqrt{\left(\varepsilon^-_{\bf k}\right)^{2}+W_0^{2}}}$.
The coupling of the OPs in the $\alpha$ and $\beta$ bands results
from momentum dependent interband Cooper pair scattering described
by terms of the form $\langle \alpha^\dagger_{{\bf k}, \uparrow}
\alpha^\dagger_{-{\bf k}, \downarrow} \rangle \beta_{{\bf k+q},
\uparrow} \beta_{-{\bf k-q}, \downarrow}$ in the MF Hamiltonian.
Below, we use $V({\bf k,k^\prime}) = V_{s}$ and $V({\bf k,k^\prime})
= V_{d} \varphi_{\bf k} \varphi_{\bf k'}/4$ with $\varphi_{\bf k} =
\cos k_{x} - \cos k_{y}$ as the pairing interactions in the $s$- and
$d_{x^2-y^2}$-wave channel, respectively. We checked that in the
limit considered here, i.e, $\Delta_{sc} \ll W_0$, feedback effects
of superconductivity on the SDW order are negligible, and moreover,
pairing terms of the form $\langle \alpha^\dagger_{{\bf k},
\uparrow} \beta^\dagger_{-{\bf k}, \downarrow} \rangle$ are
irrelevant due to the FS mismatch between the $\alpha$ and $\beta$
bands.

We begin by discussing the form of the SC OP. In the  $s$-wave case,
the SDW coherence factors entering the gap equation,
Eq.(\ref{eq:gaps}), can be factorized and one finds $\Delta_{\bf
k}^{\alpha} = D_k^\alpha \Delta_0^\alpha \left(u_{\bf k}^2-v_{\bf
k}^2\right)$ with $\Delta_0^\alpha=F^{\alpha}-F^{\beta}$, and
$\Delta_{\bf k}^{\beta}$ follows \emph{via} $\alpha \leftrightarrow
\beta$. Here, $F^{\gamma} = V_s \sum_{{\bf p}} D_{\bf p}^\gamma
\left(u_{\bf p}^2-v_{\bf p}^2\right)\f{\Delta_{\bf
p}^{\gamma}}{\Omega_{\bf p}^{\gamma}} \tanh\left(\frac{\Omega_{\bf
p}^{\gamma}}{2T}\right)$ with $\gamma = \alpha, \beta$. Moreover,
$D_{\bf k}^\gamma$ is unity if $|E_{\bf k}^{\gamma}| \leq \hbar
\omega_D$, and zero otherwise, with $\omega_D$ being the Debye
frequency. The above form of the SC gap yields three important
results. First, the $s$-wave gap is dressed by the SDW coherence
factors, and hence acquires lines nodes along the RBZ boundary where
$u_{\bf k}^2-v_{\bf k}^2=0$. Second, there is a $\pi$-phase shift
between the SC OP in the $\alpha$- and $\beta$-bands, as well as
within the same band between momenta connected by ${\bf Q}$, i.e.,
$\Delta_{\bf k}^{\gamma}=-\Delta_{\bf k+Q}^{\gamma}$. As a result,
the $s$-wave symmetry of the SC OP is \emph{unconventional}, as
summarized in Fig.\ref{fig:Fermisurf}(a). Third, while
$\Delta_0^\gamma$ needs to be calculated self-consistently, its
temperature dependence does not affect the momentum dependence of
the SC OP.

In the $d$-wave case, the SDW coherence factors can again be
factorized, and one finds from Eq.(\ref{eq:gaps}) $\Delta_{\bf
k}^{\gamma} = \varphi_{\bf k} (\Delta_0^{\gamma}+u_{\bf k}v_{\bf
k}\Delta_1^{\gamma})$ with $\Delta_0^\gamma=
F_0^{\alpha}+F_0^{\beta}$,
$\Delta_1^\alpha=-\Delta_1^\beta=F_1^{\beta}-F_1^{\alpha}$. Here,
$F_0^{\gamma} = \f{V_d}{4} \sum_{{\bf p}} D_{\bf p}^\gamma \varphi_{\bf
p}\f{\Delta_{\bf p}^{\gamma}}{\Omega_{\bf p}^{\gamma}}
\tanh\left(\frac{\Omega_{\bf p}^{\gamma}}{2T}\right)$ and
$F_1^{\gamma} = \f{V_d}{4} \sum_{{\bf p}} D_{\bf p}^\gamma \varphi_{\bf p}
u_{\bf p} v_{\bf p} \f{\Delta_{\bf p}^{\gamma}}{\Omega_{\bf
p}^{\gamma}}\tanh\left(\frac{\Omega_{\bf p}^{\gamma}}{2T}\right)$.
Since $|\Delta_1^{\gamma}|<|\Delta_0^{\gamma}|$, the SC OP in the
$\alpha$ and $\beta$-bands are in phase with no \emph{additional}
line nodes, in contrast to the $s$-wave case. The resulting phase of
the SC OP is shown in Figs.~\ref{fig:Fermisurf}(b). Note that the
temperature evolution of the self-consistently determined
$\Delta^\gamma_{0,1}$ can lead to changes in the momentum dependence
of the $d$-wave OP. Finally, we find that for both SC symmetries,
the SC OP evolves continuously into that of the paramagnetic state,
obtained in the limit $W_0 \to 0$ in which the folded parts of the
FS disappear. In particular, for the $s$-wave case, the same sign of
the SC gap is restored over the entire (large) FS.

\begin{figure}[!t]
\includegraphics[width=8.25cm]{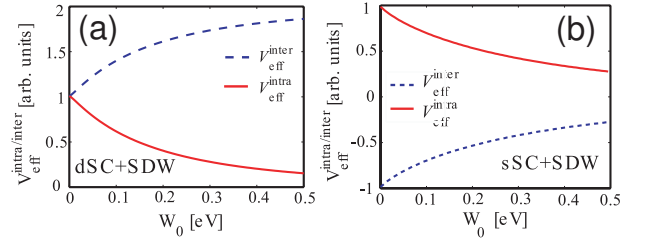}
\caption{(color online) The projected inter- and intra-band
effective interactions as a function of $W_{0}$. } \label{fig3}
\end{figure}

We next study the dependence of $T_c$ on the SDW order parameter,
$W_0$. To this end, we linearize Eq.(\ref{eq:gaps}), and solve it
iteratively in the RBZ on a $500\times 500$ lattice, setting
$\omega_D=0.1$ eV. This approach also yields the momentum dependence
of the SC gap at $T_c-0^+$. A study of the temperature induced
changes in the momentum dependence of the $d$-wave OP will be
reserved for future work. In Figs. \ref{fig2}\,(a) - (d) we present
$T_c(W_0)/T_c(W_0=0)$ as a function of $W_0$ for the $d_{x^2-y^2}$-
and $s$-wave channels and the \emph{n}- and \emph{p}-type cuprates.
In both channels, $T_c$ decreases with increasing $W_0$. For the
$d$-wave symmetry, we find that $T_c$ becomes exponentially
suppressed once $W_0$ exceeds $W_{cr1}$, where the first FS pockets
disappear in the pure SDW state. In order to understand this rapid
decrease of $T_c$ around $W_{cr1}$, we present in
Fig.~\ref{fig3}\,(a) the dependence of the effective intra- and
interband interaction projected onto the $s$- or $d$-wave channel,
defined \emph{via} $V_{eff}^{intra}=\sum_{\bf k,p}^\prime L_{\bf
k,p}^{\alpha \alpha} \phi_{\bf k} \phi_{\bf p}$ and
$V_{eff}^{inter}=\sum_{\bf k,p}^\prime L_{\bf k,p}^{\alpha \beta}
\phi_{\bf k} \phi_{\bf p}$ where $\phi_{\bf k}=1$ for the $s$-wave
case, and $\phi_{\bf k}=\cos k_x-\cos k_y $ for the $d$-wave case.
For the $d$-wave case, $V_{eff}^{intra}$ decreases with increasing
$W_0$, due to the vanishing of the SDW coherence factors in
$L^{\alpha \alpha}_{\bf k,p}$ for ${\bf k-p=Q}$, in agreement with
previous results \cite{schrieffer}. The same argument, however, does
not apply to $V_{eff}^{inter}=2-V_{eff}^{intra}$ which increases
with increasing $W_0$, thus stabilizing the SC state. Once the FS of
one of the bands disappears at $W_{cr1}$ (independent of whether
these are the electron or hole pockets), the channel for interband
Cooper pair scattering, and hence $T_c$, become rapidly suppressed
and vanish slightly above $W_{cr1}$ due to the finite Debye
frequency. While one might expect to find an exponentially
suppressed, but non-zero $T_c$ as long as the system is metallic,
i.e., for $W_{cr1}<W_0<W_{cr2}$, its resolutions is currently beyond
our numerical capabilities. This strong suppression of $T_c$ around
$W_{cr1}$ provides an explanation for the experimental observations
that the emergence of hole pockets in \emph{n}-type cuprates
coincides approximately with the onset of superconductivity on the
underdoped side \cite{review,matsui,Qazilbash}. For the $s$-wave
case, $T_c$ decreases more rapidly than for the $d$-wave case, and
becomes smaller than our numerical resolution already for $W_0$
considerably smaller than $W_{cr1}$. This behavior does not arise
from a change in the DOS, which remains almost unchanged for
$W_0<W_{cr1}$ [see Figs.~\ref{fig2}\,(e) and (f)], but is due to a
decreasing magnitude of the effective interaction,
$V_{eff}^{intra}=-V_{eff}^{inter}$, with increasing $W_0$, as shown
in Fig.~\ref{fig3}\,(b). This decrease arises since the
contributions to Eq.(\ref{eq:gaps}) coming from the folding of the
original BZ onto the RBZ, are pairbreaking. In contrast, in the
$d$-wave case, these contributions have the same sign as those
coming from the RBZ, and thus are not pairbreaking.

Finally, in Fig.\ref{fig:Fermisurf} (c) and (d), we present the
momentum dependence of the SC gaps for both symmetries along the
electron and hole FS pockets as depicted in
Fig.~\ref{fig:Fermisurf}\,(a). In agreement with the above
discussion, the $s$-wave OP is \emph{unconventional} in that it
possesses a $\pi$-phase shift between the $\alpha$- and
$\beta$-bands, and lines nodes along the RBZ boundary. In the
$d_{x^2-y^2}$-wave case, the gap is nodeless on the $\alpha$-band
and possess a line node in the $\beta$-band. While both the
\emph{unconventional} $s$- as well as the $d$-wave SC OP possess
line nodes the slope with which the gaps increase as one moves away
from the line nodes is considerably larger in the $s$-wave case,
where it is determined by the Fermi velocity, than in the $d$-wave
case, where it is set by the gap velocity [see
Fig.~\ref{fig:Fermisurf} (c) and (d)]. As a result, the DOS in both
cases scales linearly at small energies, however, with a much
smaller slope in the $s$-wave than in the $d$-wave case. In
thermodynamic measurements, it is therefore difficult to distinguish
between the \emph{unconventional} $s$-wave OP discussed above, and a
conventional, constant $s$-wave OP. We stress that the inclusion of
the SDW coherence factors in the gap equation is crucial in
determining the $W_0$-dependence of $T_c$, and momentum dependence
of the SC OP. Our results therefore lead to new properties of the
coexistence phase, not discussed in earlier studies \cite{das,ting}.

\begin{figure}[!t]
\includegraphics[width=4.5cm]{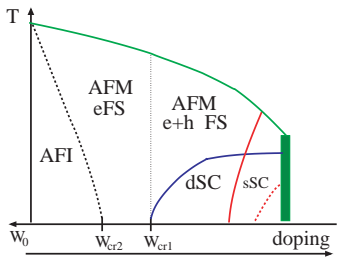}
\caption{(color online) Schematic phase diagram of the $n$-type
cuprates. The system is an antiferromagnetic metal with electron
pockets for $W_{cr1}<W_0<W_{cr2}$ (e-FS), and with electron and hole
pockets for $W_0<W_{cr1}$ (e+h-FS).} \label{fig4}
\end{figure}

The synopsis of our results, applied to the \emph{n}-type cuprates,
is shown in form of a schematic phase diagram in Fig.~\ref{fig4}.
The system is an antiferromagnetic insulator for $W_0>W_{cr2}$, and
becomes metallic with the emergence of electron pockets in the
$\alpha$-band at $W_0=W_{cr2}$. While single-band superconductivity
may occur for $W_{cr1}<W_0<W_{cr2}$, the interaction necessary to
achieve any appreciable $T_c$ is necessarily large rendering this
possibility highly unlikely for both SC symmetries in real systems.
However, the appearance of hole pockets in the $\beta$-band at
$W_{cr1}$, allows for the emergence of two-band superconductivity
with $d$-wave symmetry (blue curve). Upon further decreasing $W_0$,
the current experimental results suggest one of two scenarios.
First, superconductivity with an $s$-wave symmetry remains the
subdominant OP, implying $T^{s}_{c0}<T^{d}_{c0}$ even for $W_0
\rightarrow 0$ (red dashed curve). Second, if, as suggest by
experiments, there is a transition of the SC symmetry from $d$- to
$s$-wave (red solid curve), then this would naturally occur with
decreasing $W_0$ if for $W_0 \rightarrow 0$ one has
$T^{s}_{c0}>T^{d}_{c0}$.
The experimentally observed transition in the \emph{n}-type cuprates
between a state with well defined nodes to a state with large gapped
phase space is consistent with this second scenario. Moreover, the
observation of pair-breaking peaks in electronic Raman scattering
which occur at similar energies in both B$_{1g}$ and B$_{2g}$
channels \cite{Qazilbash}, would also suggest similar gap
magnitudes, consistent with the form of the SC gap in the $s$-wave
case (see Fig.~\ref{fig:Fermisurf}\,(c)). Clearly, further, in
particular, phase sensitive experiments are required to clarify the
SC symmetry over the entire doping range.

In conclusion, we have studied the emergence of multiband
superconductivity with $s$- and $d-$wave symmetries in the presence
of an SDW state. In both cases, the momentum dependence of the SC OP
is strongly renormalized by the SDW coherence factors. In the
$s$-wave case, this leads to an \emph{unconventional} OP which
acquires line nodes along the boundary of the RBZ, and a $\pi$ phase
shift between the $\alpha$ and $\beta$ bands. For both symmetries,
interband Cooper pair scattering stabilizes the superconducting
order. Moreover, with increasing $W_0$, the SC state with $s$-wave
symmetry is more strongly suppressed than that with $d$-wave
symmetry.  Our results are universally applicable to any (quasi-)
two-dimensional system with commensurate SDW order in which the
folding of the original BZ leads to separate FS pockets in both
bands. Finally, we note that our results can be straightforwardly
generalized to the coexistence of superconductivity with a
commensurate charge or orbital density wave state \cite{Ismer09}.

We thank A.V. Chubukov, M. Vavilov, and P. Grigoriev for the
fruitful discussions. This work is supported by the Volkswagen
Foundation (I/82203), NSF-DMR (0645461), the RMES Program (N
2.1.1/2985) (I.E.), and by the U.S. Department of Energy under Award
No. DE-FG02-05ER46225 (D.M.). D.M. and G.B. would like to
acknowledge the hospitality of the MPI-PKS where the final stages of
this manuscript were completed.

%\noindent {\small $^*$ Current address: Condensed Matter Theory
%Center, Department of Physics, University of Maryland, College Park,
%MD 20742, USA.}

\end{document}